\documentclass[a4paper]{jpconf}
\usepackage{graphicx}
\usepackage{amsmath}

\begin{document}
\title{Spontaneous Broken Local
Conformal Symmetry and Dark Energy Candidate}

\author{Lu-Xin Liu}

\address{Department of Physics, Purdue University,
West Lafayette, IN 47907, USA \\
and National Institute for Theoretical Physics, Department of Physics and Centre for Theoretical Physics,
University of the Witwatersrand, Wits, 2050, South Africa}

\ead{luxin.liu9@gmail.com}

\begin{abstract}
The local conformal symmetry is spontaneously broken down to the Local Lorentz 
invariance symmetry through the approach of nonlinear realization. The resulting 
effective Lagrangian, in the unitary gauge, describes a cosmological vector field 
non-minimally coupling to the gravitational field. As a result of the Higgs mechanism, 
the vector field absorbs the dilaton and becomes massive, but with an independent energy 
scale. The Proca type vector field can be modelled as dark energy candidate. The possibility that it further triggers Lorentz symmetry 
violation is also pointed out.   
\end{abstract}

\section{Introduction}
Astronomical observations have shown strong evidence for the existence of the dark energy, which is assumed to be 
smoothly distributed through the universe. It accounts for about 70\% of the total mass energy of the universe and 
tends to increase the expanding rate of the universe. In practice, the candidate for the dark energy, besides the cosmological 
constant, is usually taken to be a dynamical component [1]. Recently, a massive vector field as dark energy candidate has been attracted much attention [2-3]. There, 
the dark energy is modeled as a massive cosmological vector field, coupling to the gravity non-minimally and permeating 
the whole universe. 
Although the vector field can be regarded as an origin for dark energy and drives the inflation of the universe, 
the origin of the vector field itself may have different physical motivations. For example, as pointed out in Ref.[4], 
a vector field was introduced as the requirement of local scale invariance. However, one may note that, 
for a realistic physical model, there is no scale symmetry. Since at low energy the vacuum is invariant 
under (local) Lorentz transformations, it is conceivable that at higher energy matter might transform under 
a larger spacetime group. Therefore, if the scale symmetry is realized in nature, if must be a broken one.

     Motivated by that, we consider the case that the full symmetry represented by the conformal group, 
which contains the Lorentz group as a subgroup. Since the breaking mechanism which maintains 
the preponderance of the dynamical constraints of the symmetry and hence is both theoretically 
and phenomenologically most attractive is a spontaneous one, in the present paper we consider the
 spontaneous breaking of conformal symmetry and illustrate an alternative mechanism of introducing a 
massive vector field into the theory.

    As for the spontaneous symmetry breaking, the approach of nonlinear realization has been 
demonstrated a natural, economical and elegant framework for treating it. This method has been 
applied to a wide range of physical problems most notably in the form of nonlinear sigma models, 
supersymmetry and brane theories [5,6]. There, the Lagrangian is invariant with respect to 
the transformations of some continuous group $G$, but the ground state is not an invariant of $G$ but 
only of some subgroup $H$[7]. In this context, the resulting phenomenological Lagrangian becomes 
an effective theory at energies far below the scale of spontaneous symmetry breaking. If one takes 
the Lorentz group as the stability group $H$, the conformal symmetry can then be globally spontaneously 
broken down through the approach of nonlinear realization. Consequently, the effective action is 
expressed in terms of the dynamics of a Nambu-Goldstone field, i.e. the dilaton [8, 9].

    Alternatively, the conformal symmetry can also be spontaneously broken down locally. In particle 
physics, it is well known that the �compensating� gauge field need be introduced for the local gauge 
invariance of the theory. It is our purpose of this paper to point out a viable original of the massive 
cosmological vector field, as well as to give a natural explanation of this vector field at both theoretical 
and phenomenological levels. It is shown that the vector field behaves as the compensating field resulting 
from the spontaneous breaking of local conformal invariance. Furthermore, due to the Higgs mechanism, the 
vector field absorbs the dilaton particle and becomes massive. However, it has an independent mass scale. 
As a result, the massive vector particle is taken as the source of the dark energy, composing the cosmological 
gas and driving the universe�s accelerated expansion. Its mass scale is fixed by the cosmological observation.

    The paper is organized as follows. In section 2, we start by giving a brief review of the relevant spontaneous 
breaking of the global conformal symmetry. The effective action is shown to have a very natural interpretation 
in terms of Goldstone particles for broken symmetries. Then we generalize the nonlinear realization 
approach to the spontaneous breaking of the local conformal symmetry. In order to conform to the covariant 
transformations of the local Maurer-Cartan one-forms, the �compensating� gauge fields are introduced. Their 
transformations are then derived accordingly. In section 3, the effective action of the theory is constructed. 
By using the fact that the inhomogeneous transformations of the Nambu-Goldstone fields are spacetime dependent, 
we have the degree of freedom of fixing the gauge. It is shown, under the unitary gauge, the NG fields are 
transformed away and the effective action is reformulated into a vector-tensor theory. It is prominent 
to note that the resulting massive vector field has an independent energy scale, which is not determined by the 
conformal symmetry breaking scale.  The effective action that describing the general coupling of the vector field to the gravitational field is then obtained. Finally, we further our discussions by pointing out 
the possibility that the Lorentz invariance violation may be triggered by the spontaneous breaking of local conformal 
invariance. We hope this work may also shed light on the underlying theories related to Lorentz
 violation due to the presence of a vector field.

\section{Nonlinear Realization of Conformal Symmetry}
The global conformal symmetry breaking was discussed in [8, 9]. Here, before considering the case of 
local conformal symmetry breaking, for completeness we give a brief illustration about the dynamics resulting 
from its global symmetry breaking. The conformal group $G$ includes Poincare group as its subgroup along with the 
conformal boosts $K_\mu$ and dilatation $D$ symmetries. Choose the Lorentz group as the stability group $H$. 
It follows that the conformal symmetry 
can be spontaneously broken down to the $ISO(1,3)$ symmetry through the nonlinear realization approach. The 
Coset representative elements  
$
\Omega  = G/H
$                                                                                                                         
can be parameterized as                                                    
$$
\Omega  = e^{ix^\mu  P_\mu  } e^{i\phi D} e^{i\phi ^\mu  K_\mu  }  \\ \eqno{(1)}
$$                                                                                                            
where $x^\mu$ parameterize $1+3$ spacetime coordinates, and $\phi$,$\phi ^\mu$
correspond to the Nambu-Goldstone (NG) fields 
associated with the broken generators $D$ and $K^\mu$. The effective action describing the dynamics of these 
NG fields can be constructed from the Maurer-Cartan one-forms $ \Omega ^{ - 1} d\Omega$, 
which can be expanded with respect to the generators of the conformal group as 
\begin{align*}
\Omega ^{ - 1} d\Omega  = & i\omega ^\mu  P_\mu   + i\omega _D D + i\omega _k^\mu  K_\mu   + i\omega _M^{\mu \nu } M_{\mu \nu } =  ie^{ - \phi } dx^\mu  P_\mu   + (id\phi  - ie^{ - \phi } 2\phi ^\mu  dx_\mu  )D + \\
& ie^{ - \phi } 2\phi ^\nu  dx^\mu  M_{\mu \nu }  +   (ie^{ - \phi } 2\phi _\mu  dx^\mu  \phi ^\nu   - ie^{ - \phi } \phi ^\mu  \phi _\mu  dx^\nu   + id\phi ^\nu   - id\phi \phi ^\nu  )K_\nu                                         
\tag{2}
\end{align*}                                                                                                                 
The covariant coordinate differentials are found to be   
$
\omega ^\mu   = e_\nu  ^{\hspace{3pt}\mu}  dx^\nu   = e^{ - \phi } dx^\mu,
$                                                                                                       
in which $e_\nu  ^{\hspace{3pt}\mu}   = e^{ - \phi } \delta _\nu  ^{\hspace{3pt}\mu}$ is the vierbein. And the covariant 
derivative of the boost vector field $\phi ^\mu$ is given by the one-forms     
$
 \omega _k^\mu   =  \omega ^\nu  D_\nu  \phi ^\mu  =  \omega ^\nu  (2\phi _\nu  \phi ^\mu   - \phi ^{\mu '} \phi _{\mu '} \delta _\nu ^\mu   + e^\phi  \partial _\nu  \phi ^\mu   - e^\phi  \partial _\nu  \phi \phi ^\mu  ).                                               
$
It follows that the kinetic term and the potential term of the dilaton field have the form 
$$
S_{dilaton}  = \int {d^4 x\det eD_\mu  \phi ^\mu  } +  \int {d^4 } x\det e= \frac{1}{2}\int {d^4 } xe^{ - 2\phi } \partial _\mu  \phi \partial ^\mu  \phi +  \int {d^4 } xe^{ - 4\phi } \\ \eqno{(3)}
$$                                                                      
Besides, the role of connection is played by the one-forms $\omega _{M\rho }^{\mu \nu }  $ in Eq.(2), i.e.            
$$
\omega _M^{\mu \nu }  = dx^\rho  \omega _{M\rho }^{\mu \nu }  = e^{ - \phi } 2\phi ^\nu  dx^\mu   = dx^\rho  2e^{ - \phi } \delta _\rho  ^{\hspace{3pt}\mu}  \phi ^\nu   \\ \eqno{(4)} 
$$
namely,
$
 \omega _{M\rho }^{\mu \nu }  = 2e^{ - \phi } \delta _\rho  ^{\hspace{3pt}\mu}  \phi ^\nu.
$                                                                                                             
Therefore the effective action of the scalar field $\Phi$ coupling to the dilaton can be written as
\begin{align*}
S \propto & \int {d^4 x\det e\eta ^{\sigma \lambda } e_\sigma  ^{ - 1\mu } e_\lambda  ^{ - 1\nu } (\partial _\mu   + i\omega _{M\mu }^{\alpha \beta } (M_{\alpha \beta } ))\Phi } (\partial _\nu   + i\omega _{M\nu }^{\alpha \beta } (M_{\alpha \beta } ))\Phi  \\ 
  =& \int {d^4 x\det e\eta ^{\sigma \lambda } e_\sigma  ^{ - 1\mu } e_\lambda  ^{ - 1\nu } \partial _\mu  \Phi } \partial _\nu  \Phi  =  \int {d^4 x\eta ^{\mu \nu } e^{ - 2\phi } \partial _\mu  \Phi } \partial _\nu  \Phi                               
\tag{5}
\end{align*} 
For the case of fermion field, the effective action can be derived in a similar approach, 
with $M_{\alpha \beta}$ replaced by the corresponding representation of the fermion field for the Lorentz group instead.

Consider the global left action of the full group elements $g$ on the Coset elements 
Eq.(1). The result can be uniquely decomposed as the product of the new Coset elements $\Omega '$ and 
the stability group elements $h$, i.e.
$$
g\Omega  = \Omega 'h \\ \eqno{(6)}
$$                                                                                                   
in which the parameters of the full group elements
$
g = e^{ia^\mu  P_\mu  } e^{icD} e^{ib^\mu  K_\mu  } e^{i\theta ^{\mu \nu } M_{\mu \nu } }
$                                                                                                
are spacetime independent. As a result, the transformations of the Cartan one-forms become
\begin{align*}
\Omega '^{ - 1} d\Omega ' = &(h\Omega ^{ - 1} g^{ - 1} )d(g\Omega h^{ - 1} )  =   h(\Omega ^{ - 1} d\Omega )h^{ - 1}  + hdh^{ - 1}                                                 
\tag{7}
\end{align*}                                                                                                    
or namely
$
i\omega '(x') = hi\omega (x)h^{ - 1}  + hdh^{ - 1}.
$                                                                                          
Therefore, $\omega ^\mu  ,\omega _D  $ and $\omega _k^\mu$ transform covariantly under the action of Eq.(6), 
whereas $\omega _M^{\mu \nu }$ transforms by a shift $hdh^{ - 1}$ as required by the connection. On the other hand, the spontaneous breaking of 
the symmetry group can be spactetime dependent [10,11,12]. In the present context, the transformation of Eq.(6) 
becomes a local one, i.e.
$$
g = e^{ia^\mu  (x)P_\mu  } e^{ic(x)D} e^{ib^\mu  (x)K_\mu  } e^{i\theta ^{\mu \nu } (x)M_{\mu \nu } }  \\ \eqno{(8)}
$$                                                                                
in which the parameters of the group elements $g$ become spacetime dependent. 
Accordingly, the Cartan one-forms transform as
\begin{align*}
\Omega '^{ - 1} d\Omega ' = &(h\Omega ^{ - 1} g^{ - 1} (x))d(g(x)\Omega h^{ - 1} )
=  h\Omega ^{ - 1} (g^{ - 1} (x)dg(x))\Omega h^{ - 1}  + h(\Omega ^{ - 1} d\Omega )h^{ - 1}  + hdh^{ - 1}  \\ 
  \ne & h(\Omega ^{ - 1} d\Omega )h^{ - 1}  + hdh^{ - 1}  
\tag{9}
\end{align*}                                                                                                                                                                                           
Introducing the compensating one-forms into Eq.(2), one has 
$$
\Omega ^{ - 1} d\Omega  \to \Omega ^{ - 1} (d + i\hat E)\Omega = i\omega ^m P_m  + i\omega _D D + i\omega _k^m K_m  + i\omega _M^{mn} M_{mn}
  \\ \eqno{(10)}
$$                                                                                                 
Under the action of Eqs.(6,8),  it follows that the (local) Cartan one-forms transform  covariantly 
 \begin{align*}
\Omega ^{ - 1} (d + i\hat E)\Omega  \to & \Omega '^{ - 1} (d + i\hat E')\Omega ' = h(\Omega ^{ - 1} (d + i\hat E)\Omega )h^{ - 1}  + hdh^{ - 1}  
\tag{11}
\end{align*}    
It therefore leads to the infinitesimal transformations of the compensating one-forms                                                               
$$
\hat E' = g(x)\hat Eg(x)^{ - 1}  - ig(x)dg(x)^{ - 1}  \\ \eqno{(12)}
$$                                                                                    
In terms of its expanding
$
\hat E = \hat E^m P_m  + \hat AD + \hat B^m K_m  + \hat m^{mn} M_{mn},
$                                                                              
the transformations of its components are found to be
 \begin{align*}
\hat E'^m  = & \hat E^m  + 2\hat E_{m'} \theta ^{m'm} (x) + \hat E^m c(x) - \hat Aa^m (x) + 2a_n (x)\hat m^{mn}  - da^m (x) \\ 
 \hat A' =& \hat A + 2\hat E^m b_m (x) - 2a^m (x)\hat B_m  - dc(x) \\ 
 \hat B'^m  =& \hat B^m  + b^m (x)\hat A + 2\theta ^{nm} (x)\hat B_n  + 2b_n (x)\hat m^{mn}  - c(x)\hat B^m  - db^m (x)e^{ - c(x)}  \\ 
 \hat m'^{mn}  =& \hat m^{mn}  - 2\hat E^m b^n (x) + 2a^m (x)\hat B^n  - \theta ^{m'm} (x)\hat m^n _{\hspace{3pt}m'}  - \theta ^{mn'} (x)\hat m_{n'} ^{\hspace{3pt}n}  \\ 
 & + \theta _{n'} ^{\hspace{3pt}m} (x)\hat m^{n'n}  + \theta ^m _{\hspace{3pt}n'} (x)\hat m^{nn'}  - d\theta ^{mn} (x) 
\tag{13}
\end{align*}                                                                                                                     
Estimating the expansion of local Cartan one-forms of Eq.(10) with the use of the translated form of
$
\hat E = e^{ix^\mu  P_\mu  } Ee^{ - ix^\mu  P_\mu  } 
$                                                                                                  
yields 
\begin{align*}
\omega ^m  = & dx^\mu  e^{ - \phi } (E_\mu  ^{\hspace{3pt}m}  + \delta _\mu  ^{\hspace{3pt}m} ) = dx^\mu  e_\mu  ^{\hspace{3pt}m}  \\ 
 \omega _D  = & dx^\mu  ( - 2E_\mu  ^{\hspace{3pt}m} \phi _m e^{ - \phi }  + A_\mu   + \partial _\mu  \phi  - 2e^{ - \phi } \phi _m \delta _\mu  ^{\hspace{3pt}m} ) = dx^\mu  \omega _{D\mu }  \\ 
 \omega _k^m  = & dx^\mu  (2\phi ^m \phi _n E_\mu  ^{\hspace{3pt}n} e^{ - \phi }  - \phi _n \phi ^n E_\mu  ^{\hspace{3pt}m} e^{ - \phi }  - \phi ^m A_\mu   + B_\mu  ^{\hspace{3pt}m} e^\phi   + 2\phi _n m_\mu  ^{\hspace{3pt}nm}  +  \\ 
 & \partial _\mu  \phi ^m  - \partial _\mu  \phi \phi ^m  + 2e^{ - \phi } \phi ^m \phi _n \delta _\mu  ^{\hspace{3pt}n}  - e^{ - \phi } \phi _n \phi ^n \delta _\mu  ^{\hspace{3pt}m} )  =  dx^\mu  \omega _{k\mu }^m  \\ 
 \omega _M^{mn}  = & dx^\mu  (2\phi ^n E_\mu  ^{\hspace{3pt}m} e^{ - \phi }  + m_\mu  ^{\hspace{3pt}mn}  + 2e^{ - \phi } \phi ^n \delta _\mu  ^{\hspace{3pt}m} ) = dx^\mu  \omega _{M\mu }^{mn}  
\tag{14}
\end{align*}                                                                                         
where the Latin indices $m,n,...$ represent the local Lorentz indices, and the Greek indices $ \mu ,\nu ,...$ represent 
the general coordinate indices. According to Eq.(11), their infinitesimal transformation laws are found to be
\begin{align*}
\omega '^m  = & \omega ^n L_n ^{\hspace{3pt}m};
 \omega _D ' = \omega _D;
 \omega _k '^m  = \omega _k^n L_n ^{\hspace{3pt}m};
 \omega _M '^{mn}  = \omega _M^{ab} L_a ^{\hspace{3pt}m} L_b ^{\hspace{3pt}n}  - d\lambda ^{mn} (x) 
\tag{15}
\end{align*}                                                                                                                                       
In terms of their components, Eq.(15) can be reexpressed as
$$
{e'}_{\mu}^{\hspace{3pt}m}  = \frac{{\partial x^\nu  }}{{\partial {x'}^\mu  }}L_n ^{\hspace{3pt}m} e_{\nu}^{\hspace{3pt}n};
 {\omega '}_{D\mu }  = \frac{{\partial x^\nu  }}{{\partial {x'}^\mu  }}\omega _{D\nu }; 
 {\omega '}_{k\mu }^m  = \frac{{\partial x^\nu  }}{{\partial {x'}^\mu  }}L_n ^{\hspace{3pt}m} \omega _k^n  \\ \eqno{(16)}
$$
along with the inhomogeneous transformations of the spin connection
$
 {\omega '}_{M\mu} ^{mn}  = \frac{{\partial x^\nu  }}{{\partial x'^\mu  }}(\omega _{M\nu} ^{ab} L_a ^{\hspace{3pt}m} L_b ^{\hspace{3pt}n}  - \partial _\nu  \lambda ^{mn} (x)),	
$                                                                             
where $L_n ^{\hspace{3pt}m}$ is the spacetime dependent matrix
 related to the local Lorentz transformation $h$, which is parameterized as $ h = e^{i\lambda ^{\mu \nu } (x)M_{\mu \nu } }$,
 and $\lambda ^{\mu \nu } (x)$ is a function of spacetime determined by the relation of Eq.(6). 

\section{Effective Actions}
The transformation properties of these Cartan one-forms are secured by the transformations (11), 
and this, in conjunction with the requirement that the phenomenological Lagrangian be invariant under 
these transformations, describes the underlying theory whose symmetry structure becomes a dynamical type. 
In fact, the effective actions, which are locally conformal invariant, can be constructed by using the
 building blocks of the one-forms Eq.(16) as well as their covariant derivatives. Therefore, according to Eq.(16) the rank 
two metric tensor
$
g_{\mu \nu }  = \eta _{mn} e_\mu  ^{\hspace{3pt}m} e_\nu  ^{\hspace{3pt}n} $  transforms as 
\begin{align*}
{g'}_{\mu \nu }  = & \eta _{mn} {e'}_\mu  ^{\hspace{3pt}m} {e'}_\nu  ^{\hspace{3pt}n}  = \eta _{mn} \frac{{\partial x^\gamma  }}{{\partial {x'}^\mu  }}L_a ^{\hspace{3pt}m} e_\gamma  ^{\hspace{3pt}a} \frac{{\partial x^{\nu '} }}{{\partial {x'}^\nu  }}L_{n'} ^{\hspace{3pt}n} e_{\nu '} ^{\hspace{3pt}n'} 
  =  \frac{{\partial x^\gamma  }}{{\partial {x'}^\mu  }}\frac{{\partial x^{\nu '} }}{{\partial {x'}^\nu  }}\eta _{an'} e_\gamma  ^{\hspace{3pt}a} e_{\nu '} ^{\hspace{3pt}n'}  = \frac{{\partial x^\gamma  }}{{\partial x'^\mu  }}\frac{{\partial x^{\nu '} }}{{\partial {x'}^\nu  }}g_{\gamma {\nu '}},  
\tag{17}
\end{align*} 
and the vierbein has an explicit form
$
e_\mu  ^{\hspace{3pt}m}  = e^{ - \phi } (E_\mu  ^{\hspace{3pt}m}  + \delta _\mu  ^{\hspace{3pt}m} ).
$                               

The metric tensor $g_{\mu \nu }$ and its inverse $g^{\mu \nu }$ are used 
to raise and lower indices on world tensors. The covariant derivative of the local 
Lorentz tensor $T^a _{\hspace{3pt}b}$ is defined as
$$
\nabla _\nu  T^a _{\hspace{3pt}b}  = \partial _\nu  T^a _{\hspace{3pt}b}  - \omega _{M\nu b}^c T^a _{\hspace{3pt}c}  + \omega _{M\nu c}^a T^c _{\hspace{3pt}b}  \\ 
 \eqno{(18)}
$$                                                                            
in which the spin connection $\omega _{M\nu }^{ab}$ is given by Eq.(14), whereas
 the covariant derivative of mixed tensor $T_\mu  ^{\hspace{3pt}m}$ takes the form
$$
\nabla _\nu  T_\mu  ^{\hspace{3pt}m}  = \partial _\nu  T_\mu  ^{\hspace{3pt}m}  - \Gamma _{\nu \mu }^\lambda  T_\lambda  ^{\hspace{3pt}m}  + \omega _{M\nu n}^m T_\mu  ^{\hspace{3pt}n}  \\ 
\eqno{(19)}
$$                                                                             
The local Lorentz tensor $T^{mn}$ is transcribed 
into a world tensor $T^{m\nu }$ by means of the inverse of 
vierbein in the usual way, i.e. $T^{m\nu }  = e^{ - 1\nu } _{\hspace{9pt}n} T^{mn}$, and vice versa. 
And the affine connection $\Gamma _{\nu \mu }^\lambda$ is given by 
$$
\Gamma _{\mu \nu }^\rho   = e^{ - 1\rho } _{\hspace{9pt}a} \partial _\mu  e_\nu  ^{\hspace{3pt}a}  + e^{ - 1\rho } _{\hspace{9pt}a} e_\nu  ^{\hspace{3pt}b} \omega _{M\mu }^{ac} \eta _{cb}  
\\
\eqno{(20)}
$$                                                                               
which can be derived from the requirement that the covariant derivative of the vierbein vanishes
\begin{align*}
 \nabla _\nu  e_\mu  ^{\hspace{3pt}m}  = & \partial _\nu  e_\mu  ^{\hspace{3pt}m}  - \Gamma _{\nu \mu }^\lambda  e_\lambda  ^{\hspace{3pt}m}  + \omega _{M\nu n}^m e_\mu  ^{\hspace{3pt}n}=0
\tag{21}
\end{align*}                                                                                                                                   
Reinterpreting Eq.(21) in terms of the metric tensor $g^{\mu \nu }$, one finds 
$
\nabla _\rho  g^{\mu \nu }  = 0 = \partial _\rho  g^{\mu \nu }  + \Gamma _{\sigma \rho }^\mu  g^{\sigma \nu }  + \Gamma _{\sigma \rho }^\nu.    
$                                                                        
This, in conjunction with the torsion free condition
\begin{align*}
\Gamma _{\mu \nu }^\rho   - &\Gamma _{\nu \mu }^\rho   = e^{ - 1\rho } _{\hspace{9pt}a} (\partial _\mu  e_\nu  ^{\hspace{3pt}a}  + e_\nu  ^{\hspace{3pt}b} \omega _{M\mu }^{ac} \eta _{cb}  - \partial _\nu  e_\mu  ^{\hspace{3pt}a}  + e^{ - 1\rho } _{\hspace{9pt}a} e_\mu  ^{\hspace{3pt}b} \omega _{M\nu }^{ac} \eta _{cb} )=0
\tag{22}
\end{align*}                                                                                                                                   
leads to the solution for the affine connection
$                                                         
 \Gamma _{\mu \nu }^\rho   = \frac{1}{2}g^{\rho \alpha } (\partial _\mu  g_{\nu \alpha }  + \partial _\nu  g_{\mu \alpha }  - \partial _\alpha  g_{\mu \nu } ).$ It follows that the rank four Riemann curvature tensor can be written as
$$
R^\rho  _{\hspace{3pt}\sigma \mu \nu }  = \partial _\mu  \Gamma _{\nu \sigma }^\rho   - \partial _\nu  \Gamma _{\mu \sigma }^\rho   + \Gamma _{\nu \sigma }^\alpha  \Gamma _{\mu \alpha }^\rho   - \Gamma _{\mu \sigma }^\beta  \Gamma _{\nu \alpha }^\rho   \\ 
\eqno{(23)}
$$                                                                      
By means of contracting the Riemann curvature tensor, the associated Ricci tensor is defined as 
$
R_{\mu \nu }  = R^\alpha  _{\hspace{3pt}\nu \mu \alpha }, 
$
whereas the curvature scalar 
$
R = g^{\mu \nu } R_{\mu \nu }$                                                                                                                  
is defined as the trace of Ricci tensor. By using the affine connection $\Gamma _{\nu \mu }^\lambda$, 
the covariant derivative of the one-forms $\omega _{D\mu }$ is written as
$$
\nabla _\nu  \omega _{D\mu }  = \partial _\nu  \omega _{D\mu }  - \Gamma _{\nu \mu }^\alpha  \omega _{D\alpha }  \\ 
\eqno{(24)}  
$$
Considering Eqs.(16,17), the quadratic term for one-forms $\omega _{D\mu }$ is locally conformal invariant
$$
\frac{1}{2}g_{\mu \nu } \omega _D ^\mu  \omega _D ^\nu  M^2  = \frac{1}{2}\omega _{D\mu } \omega _D ^\mu  M^2  \\ 
\eqno{(25)}  
$$
this will lead to the mass term of the vector field after fixing the gauge of 
the conformal transformations. Besides, the kinetic terms of $\omega _{D\mu }$ containing no higher 
than second derivatives have the form
$$
 - \frac{1}{2}\nabla _\nu  \omega _{D\mu } \nabla _\alpha  \omega _{D\beta } (g^{\nu \alpha } g^{\mu \beta }  - g^{\nu \beta } g^{\mu \alpha } ) \\ 
\eqno{(26)}
$$
as well as 
$$
\nabla _\nu  \omega _{D\mu } \nabla ^\nu  \omega _D ^\mu   \\ 
\eqno{(27)}
$$                                                                                                                  
Explicitly, both of them are locally conformal transformation invariant.

     Since the vierbein transforms as ${e'}_\mu  ^{\hspace{3pt}m}  = \frac{{\partial x^\nu  }}{{\partial {x'}^\mu  }}L_n ^{\hspace{3pt}m} e_\nu  ^{\hspace{3pt}n}$, 
we have            
\begin{align*}
 d^4 x'\det e' = & d^4 x\det |\frac{{\partial x'^\nu  }}{{\partial x^\mu  }}|\det |\frac{{\partial x^\mu  }}{{\partial x'^\nu  }}|\det e\det L_n ^{\hspace{3pt}m} =  d^4 x\det e\det L_n ^{\hspace{3pt}m}  = d^4 x\det e 
\tag{28}
\end{align*}                                                                                                                                                                
This, along with Eqs.(24-27), leads to the general locally conformal invariant effective action, which takes the form
\begin{align*}
 I = & \int {d^4 x\det e(}  - \frac{1}{2}\nabla _\nu  \omega _{D\mu } \nabla _\alpha  \omega _{D\beta } (g^{\nu \alpha } g^{\mu \beta }  - g^{\nu \beta } g^{\mu \alpha } ) + \varepsilon \nabla _\nu  \omega _{D\mu } \nabla ^\nu  \omega _D ^\mu   \\ 
  & + \frac{1}{2}\omega _{D\mu } \omega _D ^\mu  M^2 + \frac{1}{{16\pi G}}R + \xi \omega _{D\mu } \omega _D ^\mu  R + \eta \omega _{D\mu } \omega _{D\nu } R^{\mu \nu } ) 
\tag{29}
\end{align*}

   The action (29) is locally conformal transformation invariant, whereas upon making the gauge 
choice, Eq.(29) can be further simplified and it becomes feasible to identity the remaining of 
$e_\mu  ^{\hspace{3pt}m}$ as the physical vierbein field, and the remaining of $\omega _{M\nu }^{ab}$ as the 
physical spin connection. It can be 
seen that the local conformal gauge invariance of the action prevents the Goldstone field from 
making an explicit appearance, and thereafter the action will become locally invariant under the 
linearly realized stability subgroup $H$. We will work on the unitary gauge. Consider the transformations 
of the Nambu-Goldstone fields given by Eq.(6).                                                                                                                   
By using Eqs.(1,8), it becomes                                   
$
e^{ia^\mu  (x)P_\mu  } e^{ic(x)D} e^{ib^\mu  (x)K_\mu  } e^{i\theta ^{\mu \nu } (x)M_{\mu \nu } } e^{ix^\mu  P_\mu  } e^{i\phi (x)D} e^{i\phi ^\mu  (x)K_\mu  }
 =  e^{i{x'}^\mu P_\mu  } e^{i\phi '(x')D} e^{i\phi '^\mu  (x')K_\mu  } e^{i\lambda '^{\mu \nu } (x')M_{\mu \nu } }.  
$
Applying the variation of the full group elements $g$
$$
\delta g = i(a^\mu  (x)p_\mu   + c(x)D + b^\mu  (x)K_\mu   + \theta ^{\mu \nu } (x)M_{\mu \nu } ) \\ 
 \eqno{(30)} 
$$                                                    
to it gives us
$$
 (1 + \delta g)\Omega  = (\Omega  + \delta \Omega )(1 + \delta h) \\  
\eqno{(31)} 
$$                                                                                        
It further leads to
$
\Omega ^{ - 1} \delta g\Omega  - \Omega ^{ - 1} \delta \Omega  = \delta h.
$ Then one can find 
\begin{align*}
&\Omega ^{ - 1} i(a^\mu  (x)p_\mu   + c(x)D + b^\mu  (x)K_\mu   + \theta ^{\mu \nu } (x)M_{\mu \nu } )\Omega 
 - \Omega ^{ - 1} \delta \Omega  = i\lambda ^{\mu \nu } (x)M_{\mu \nu }  
\tag{32}
\end{align*}
The infinitesimal general coordinate transformation as well as these of the Nambu-Goldstone 
fields can be derived up to the first order by comparing the coefficients of the generators 
from each side of Eq.(32). As shown below, the general coordinates and the Nambu-Goldstone fields transform as
$$
\delta x^\mu   = a^\mu  (x) + x^\mu  c(x) + 2x \cdot b(x)x^\mu   - x^2 b^\mu  (x) + 2\theta ^{\nu \mu } (x)x_\nu   \\ 
$$
$$
\delta \phi  =  c(x) + 2x^\mu  b_\mu  (x) + ...;
 \delta \phi ^\mu   =  b^\mu  (x)e^\phi   - 2\phi ^\mu  x_\mu  b^\mu  (x) + 2\phi  \cdot b(x)x^\mu   + ... 
 \eqno{(33)} 
$$
Here, $\delta \phi$ etc are the total variation of the fields, i.e. $\delta \phi  = \phi '(x') - \phi (x)$, 
and the intrinsic variation of the these fields is defined as $\delta _{in} \phi  = \phi '(x') - \phi (x') = \delta \phi  - (\phi (x') - \phi (x)) 
  = \delta \phi  - \delta x^\mu  \partial _\mu  (\phi (x') - \phi (x))$. 
The broken generators $D$ and $K^\mu$ give rise to an inhomogeneous and nonlinear transformation law for 
the Goldstone field $\phi$ and $\phi ^\mu$, and these inhomogeneous terms become functions of spacetime and signal 
the spontaneous breaking of the conformal symmetry. A further choice of the transformations of 
the fields $\phi$ and $\phi^\mu$ corresponds to a gauge choice that leaves the Lorentz group as a residual gauge 
freedom. Choosing the unitary gauge [11,12,13], i.e. $ \phi  \to \phi ' = 0$,$\phi ^\mu   \to \phi '^\mu   = 0$,
one then has
$$
\Omega  \to \Omega ' = e^{ix'^\mu  P_\mu  }  \\
\eqno{(34)} 
$$                                                                                             
Moreover, Eq.(14) becomes	
 \begin{align*}
{{\omega '}^m } = & {{dx'}^\mu}  ({{E'}_{\mu}^{\hspace{3pt}m}}  + {{\delta '}_\mu ^{\hspace{3pt}m}} ) = {{dx'}^\mu}  {{e'}_\mu^{\hspace{3pt}m}};
 {{\omega '}_D}  =  {{dx'}^\mu}  {{A'}_\mu}   = {dx'}^\mu  {\omega '}_{D\mu }  \\ 
{ {\omega '}_k^m}  =& {{dx'}^\mu}  {{B'}_\mu^{\hspace{3pt}m}}  = {dx'}^\mu { \omega '}_{k\mu }^m ;
 {{\omega '}^{mn}_M}  = {{dx'}^\mu}  {{m'}_\mu^{\hspace{3pt}mn}}  = {dx'}^\mu  {\omega '}_{M\mu} ^{mn}  
\tag{35}
\end{align*}
Since the NG fields have been transformed away by fixing the gauge, the theory is no 
longer conformal but rather local Lorentz gauge invariant. For the sake of simplicity, 
we drop the prime in what follows of this paper. Therefore, the effective action of Eq.(29) 
is then converted to a vector-tensor theory                               
 \begin{align*}
I =& \int {d^4 x\det e(}  - \frac{1}{2}\nabla _\nu  A_\mu  \nabla _\alpha  A_\beta  (g^{\nu \alpha } g^{\mu \beta }  - g^{\nu \beta } g^{\mu \alpha } ) + \varepsilon \nabla _\nu  A_\mu  \nabla ^\nu  A^\mu   +  \\ 
 &\frac{1}{2}A_\mu  A^\mu  M^2  + \frac{1}{{16\pi G}}R + \xi A_\mu  A^\mu  R + \eta A_\mu  A_\nu  R^{\mu \nu } ) \\ 
  = &\int {d^4 } x\det e( - \frac{1}{4}F_{\mu \nu } F_{\rho \sigma } g^{\mu \rho } g^{\nu \sigma }  + \varepsilon \nabla _\nu  A_\mu  \nabla ^\nu  A^\mu   + \frac{1}{2}A_\mu  A^\mu  M^2 + \\ 
 & \frac{1}{{16\pi G}}R + \xi A_\mu  A^\mu  R + \eta A_\mu  A_\nu  R^{\mu \nu } ) 
\tag{36}
\end{align*}
where the field strength $F_{\mu \nu }  = \nabla _\mu  A_\nu   - \nabla _\nu  A_\mu   = \partial _\mu  A_\nu   - 
\partial _\nu  A_\mu $, and the non-minimal coupling of the vector 
field to the Ricci scalar and to the Ricci tensor are descried by parameters $ \xi$ 
and $\eta$ respectively. As a result of the Higgs Mechanism [14], the gauge field $A_\mu$, 
associated with the broken generator $D$, absorbs the dilaton and acquires a third polarization and becomes massive. However, 
it is prominent to note that the vector field has an independent energy scale, which is totally 
different from the case of spontaneous symmetry breaking in the standard model of particle physics. 
There, the mass of the massive fields is determined by the symmetry breaking scale instead. In the 
present context, the effective action describes a scenario that the vector 
field does not decouple at certain energy scale, and it is included as a dynamical degree of freedom in the action
accounting for the removal of the Nambu-Goldstone boson from the spectrum.

Modelled as dark energy candidate, the mass scale of the massive vector-like
 field, which is non-minimally coupled to the gravitational field, 
can be constrained from observational data 
of the cosmology. The general equation of motion of the vector field can be derived from Eq.(36) as     
$$
\nabla _\nu  F^{\mu \nu }  + 2\varepsilon \nabla _\nu  \nabla ^\nu  A^\mu   - A^\mu  M^2  - 2\xi A^\mu  R - 2\eta A^\nu  R_\nu ^\mu   = 0 \\ 
\eqno{(37)} 
$$                                         
Taking different values for the parameters would lead to different 
physical models [2,15]. Because the dynamics of vectors known to occur 
in nature are described by a Maxwell term, therefore, for the Maxwell type vector theory, 
the equation of motion (37) can be converted to the following form after using 
the relation $[\nabla _\mu  ,\nabla _\nu  ]A^\alpha   = R^\alpha  _{\hspace{3pt}\beta \mu \nu } A^\beta $ and 
setting the parameter $\varepsilon$ to zero
$$
\nabla _\nu  \nabla ^\nu  A^\mu   - \nabla ^\mu  \nabla _\nu  A^\nu   + A^\mu  M^2  + 2\xi A^\mu  R + (1 + 2\eta )A^\nu  R_\nu ^\mu   = 0 \\ 
\eqno{(38)} 
$$
Hence, the Proca type vector field interacts only with the gravitational 
field and has no standard matter interactions. Accordingly, its mass scale would be 
intertwined with large distance scale effects of the universe through the Friedmann 
Equation after one applies the equation of motion of this massive cosmological vector field 
to cosmology.  Specifically, for a homogeneous and isotropic universe, one can assume the 
smoothly distributed vector field  
$
A^\mu   = (A(t),0,0,0)  
$                                                                                           
 to be a function of time with only one non-zero time component. Adopting the Robertson-Walker metric $g_{\mu \nu }  = (1, - a(t), - a(t), - a(t))$ gives us $ds^2  = dt^2  - a^2 (t)(dx^2  + dy^2  + dz^2 )$                                                                              
for the invariant interval. Therefore, the equation of motion of $A^\mu$(Eq.(38)) leads to 
the evolution equation for the scale factor $a(t)$
$$
(16\xi  + 8\eta )\frac{{\mathop a\limits^{..} }}{a} + 16\xi \frac{{\dot a^2 }}{{a^2 }} = \frac{{4M^2 }}{3} \\ 
\eqno{(39)}
$$
In the limit of large $a$, it can be shown 
$
a = \exp [H(t - t_0 )]
$[2],
where the Hubber parameter $H = (\frac{{M^2 }}{{3(8\xi  + 2\eta )}})^{\frac{1}{2}}$ is related to 
the mass of the vector particle. Hence, due to the presence of the massive cosmological vector particle, the 
Friedmann equation gives us a scenario of an inflate universe. The mass of the 
cosmological massive vector particle which may represent the main component of the universe thus plays the role of 
cosmological constant.

To summarize, in this paper we demonstrated the existence of a massive vector field, which is the result of the spontaneous broken local conformal symmetry. We illustrate the mechanism of spontaneous symmetry breaking through nonlinear realization along with its correlation with the dark energy. As a result, this paper shows a possible origin of dark energy candidate as a Proca type vector field.  As shown above, this vector field has no standard model interactions; it has an independent energy sale, permeating all over the space. Its independent energy scale can be determined by applying to specific cosmological models along with the cosmological observations. In addition, consider an alternative local Lorentz invariant quartic term
$
\frac{1}{2}\lambda (A_\mu  A^\mu  )^2   
$
being added to the effective action (36), where $\lambda$ is the coupling constant. 
If the quadratic term has a wrong sign squared mass    
$                                                                         
  - \lambda A_\mu  A^\mu  M^2
$                                                                                                                
then the Lorentz symmetry would be spontaneously broken down [16]. Explicitly, 
the potential $V = \frac{1}{2}\lambda (A_\mu  A^\mu  )^2  - \lambda A_\mu  A^\mu  M^2$ is found to 
have a minimum when the vector field satisfies the condition $A_\mu  A^\mu   = M^2$. 
Therefore, the non-zero vacuum expectation value of the vector field $A^\mu$ signals the spontaneous 
breaking of the Lorentz symmetry, and this Lorentz violating vector would play an important role in 
cosmological models. The impact of its phenomenological consequences on physics 
should be broad and need further exploration.

\end{document}